# On Classifying Sepsis Heterogeneity in the ICU: Insight Using Machine Learning

Zina Ibrahim[1,2,3*], Honghan Wu[4], Ahmed Hamoud[5], Lukas Stappen[6], Richard Dobson[1,2,3], Andrea Agarossi[7]

[1] Department of Biostatistics & Health Informatics, King's College London, London, United Kingdom
[2] Institute of Health Informatics, University College London, London, United Kingdom
[3] Health Data Research UK London, University College London, London, United Kingdom
[4] Usher Institute of Population Health Sciences and Informatics, University of Edinburgh, Edinburgh, United Kingdom
[5] Department of Renal Medicine, East and North Hertfordshire NHS Trust, Stevenage, United Kingdom
[6] Embedded Intelligence for Health Care and Wellbeing, University of Augsburg, Augsburg, Germany
, Augsburg University, Augsburg, Germany
[7] Department of Anaesthesia and Intensive Care, Luigi Sacco Hospital, Milan, Italy

## Corresponding Author

| | |
|---|---|
| **Name:** | Zina M. Ibrahim |
| **Address:** | Department of Biostatistics & Health Informatics, Institute of Psychiatry, Psychology and Neuroscience, King's College London. P.O. 80 the Social, Genetic and Developmental Psychiatry Centre, De Crespigny Park, |
| Denmark | Hill, London SE5 8AF, United Kingdom. |
| **E-mail:** | zina.ibrahim@kcl.ac.uk |
| **Tel:** | +44 (0) 758 285 9501 |

## Mesh Terms

Machine Learning, Sepsis Syndrome

**Word Count:** 2,000

**ABSTRACT**

Current machine learning models aiming to predict sepsis from Electronic Health Records (EHR) do not account for the heterogeneity of the condition, despite its emerging importance in prognosis and treatment. This work demonstrates the added value of stratifying the types of organ dysfunction observed in patients who develop sepsis in the ICU in improving the ability to recognise patients at risk of sepsis from their EHR data. Using an ICU dataset of 13,728 records, we identify clinically significant sepsis subpopulations with distinct organ dysfunction patterns. Classification experiments using Random Forest, Gradient Boost Trees and Support Vector Machines, aiming to distinguish patients who develop sepsis in the ICU from those who do not, show that features selected using sepsis subpopulations as background knowledge yield a superior performance regardless of the classification model used. Our findings can steer machine learning efforts towards more personalised models for complex conditions including sepsis.

**BACKGROUND AND SIGNIFICANCE**

Sepsis, defined by a life-threatening response to infection and potentially leading to multiple organ failure, is one of the most significant causes of worldwide morbidity and mortality[1]. Sepsis is implicated in 6 million deaths annually, with costs totaling $24 billion in the USA alone[2].

Early identification of sepsis is a crucial factor in improving outcomes[3-5]. Yet, traditional score-based screening tools lack the specificity needed to identify and elevate the care of potentially septic patients[6-10]. In response, machine learning (ML) algorithms have been developed to recognise sepsis onset from vital signs data. A select number of ML models have shown improved predictions by taking advantage of computational power and large-scale data mining[11-13], or attempting to optimise the feature set required for prediction[14-15]. Nevertheless, current ML models have shown mixed results reflecting the heterogeneity of sepsis[16-18], populations[19] and methodologies[20].

The objective of this paper is to highlight the importance of classifying the clinical heterogeneity of sepsis in enhancing our ability to anticipate onset, with focus on sepsis developed in the ICU. By analysing routine clinical data of patients who develop sepsis in the ICU, we show that: 1) the clinical presentation of sepsis is underpinned by distinct combinations of dysfunction patterns that are mostly independent of etiology, 2) these patterns exhibit associations with distinct variations in vital signs and laboratory tests obtained within 24 hours of ICU admission, 3) using the relevant vitals and tests for each pattern as features in classification experiments produces highly sensitive and specific predictions regardless of the classification algorithm used, reflecting the relevance of the features to the clinical outcome. The results advocate that future sepsis prediction ML models can be guided towards better discriminative power by reformulating the sepsis prediction task to target the recognition of the different clinical manifestations of sepsis as opposed to the classic definitions currently in use. Such task will prioritise the features used for prediction using preprocessing steps that map a patient's routinely collected clinical data to previously derived subpopulations. Although this work does not aim to devise a sepsis prediction algorithm, it advocates a methodological shift in ML sepsis prediction, supported by recent findings of reproducible clinical phenotypes of sepsis[21].

## MATERIALS AND METHODS

### Data and Preprocessing

We used the data of ICU stays between 2001 and 2012 obtained from the anonymised Medical Information Mart for Intensive Care III (MIMICIII) database[22]. We extracted ICU stays of adults scoring a Sequential Organ Failure (SOFA) severity score $\geq 2$ with neither a primary sepsis diagnosis or suspected sepsis recorded in the ICU admission notes, and further processed the data to exclude stays shorter than 24 hours (1756 records), records with incomplete administrative information (581 records), and more than 15% missing vital signs, as they can lead to inadequate imputation (3,962 records). We used pattern matching to identify paragraphs within the admission notes containing mentions of sepsis. Records with no mention of sepsis were automatically included in our cohort (7,823 records), while the extracted paragraphs of records with sepsis mentions (27,041 paragraphs) were manually validated over a 1-year period to exclude records with ambiguous mentions, or suspected or confirmed sepsis. The final dataset[1] contained 13,728 ICU stays, with 31% (4,256) of the records having the primary outcome of sepsis (ICD-9 codes 995.91, 995.92, 785.52) (1,976) in the discharge records or satisfying the 3$^{rd}$ International definition of sepsis of a life-threatening organ dysfunction (identified as a SOFA score $\geq 2$) caused by a dysregulated host response to infection, confirmed by positive cultures (2,280 records)[8].

We extracted admission details, comorbidity indices, etiology details and the pre-calculated respiratory, cardiovascular, renal, hepatic, central nervous system and coagulation subcomponents of the SOFA score (Appendix A). We calculated 63 vitals and laboratory tests aggregated over the first 24 hours of admission (Appendix B). As in similar studies[23], we extended the window forward by 24 hours for the infrequently sampled laboratory measurements to improve data completion. We imputed missing data using k-nearest neighbour (k=7).

### Experimental Design

Figure 1 shows our approach. The idea is to cluster the ICU records with confirmed sepsis using the SOFA subcomponents to uncover subpopulations with distinct organ dysfunction patterns, and then to perform feature selection on the individual clusters to identify subsets of the 63 vitals with the highest variance in each subpopulation. We compare classification performance with sepsis diagnosis as outcome using two feature sets selected from: a) the entire septic population, and b) individual subpopulations.

### Sepsis Subpopulations

We used Self-Organisation Maps (SOM)[24] and clustering to obtain a two-dimensional visualisation of the confirmed sepsis records based on organ dysfunction patterns as done in [16]. A SOM is a powerful ML model that maps highly dimensional data into a two-dimensional grid of neurons, each corresponding to records with extremely similar features. We trained a 17x17 SOM to aggregate the 4,256 records into 289 neurons, each representing 5-35 ICU stays extremely similar in organ dysfunction types. SOM parameters were chosen heuristically by minimising the number of empty nodes and balancing the number of records mapping to each node[25].

---

[1] SQL scripts for recreating the dataset using the MIMICIII database are available at: https://github.com/KHP\-Informatics/sepsis

Consistent with existing biomedical literature citing SOMs' superiority[26-30], we did not compare SOM clustering with other techniques. As our aim is to use the SOM-based similarity to discover any underlying clusters, we used a distance-based clustering, mainly hierarchical, to minimise the within-cluster variance in SOM-generated distances. We determined 4 as the optimal number of clusters by examining hierarchical clustering dendograms over 1,000 iterations. We used the R SOM package Kohonen[31] and NBClust[32] for clustering. We report summary statistics as median/interquartile range, or count/percentage, as appropriate. We compared the central tendencies of the features using the Kruskal-Wallis test, using a cut-off value of p=0.01.

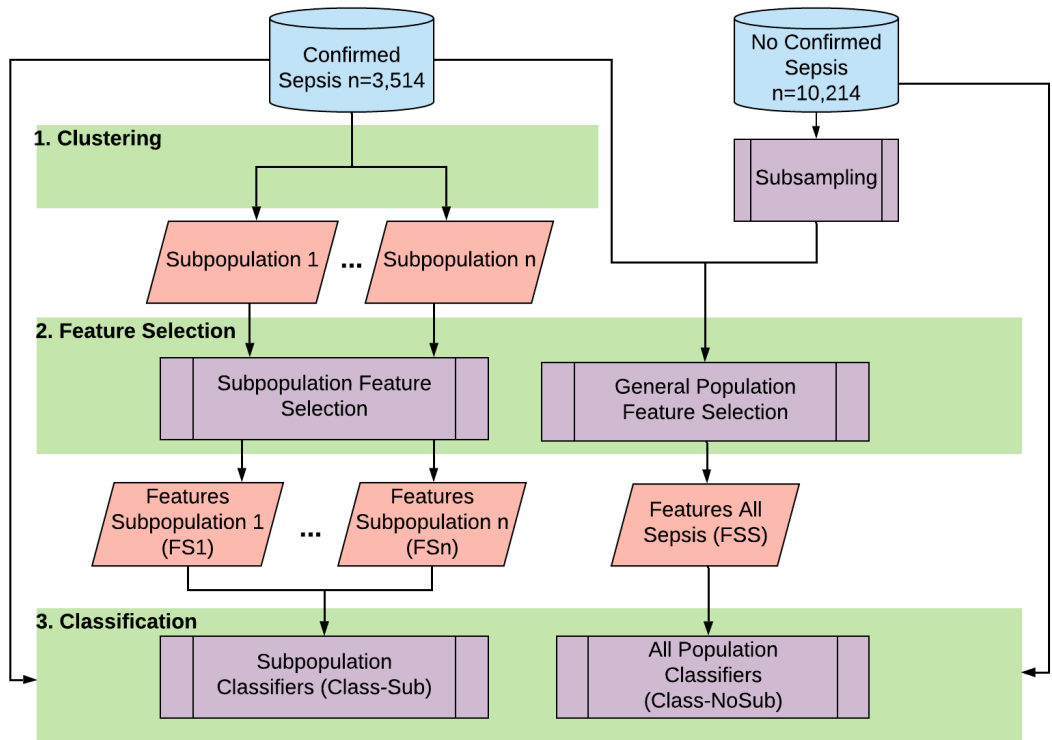

**Figure 1:** The Overall Flow of Subtype-based Sepsis Identification

**Feature Selection**
Using the 63 predictors and 4,256 records of diagnosed sepsis, we performed feature selection to identify the most relevant features for each subpopulation. We used Random Forests (RF)[33] with conditional permutation variable importance, to account for spurious correlations among predictors[34-35]. For each RF, the target cluster assignment designated positive outcomes (i.e. records with class label 1 for RF1). Additionally, we performed feature selection using all 13,728 records with sepsis diagnosis as outcome.

We trained each RF over 1,000 bootstrapped iterations to avoid overfitting, computing the importance of the 63 predictors after each. To achieve robustness against statistical fluctuations, the best features for every RF were the top quartile of a rank-invariant tally for the number of iterations a feature's importance was above the mean importance. The resulting featuresets are *FS1-FS4*, corresponding to each subpopulation, and *FSS* for the entire septic population (Figure 2 a-e).

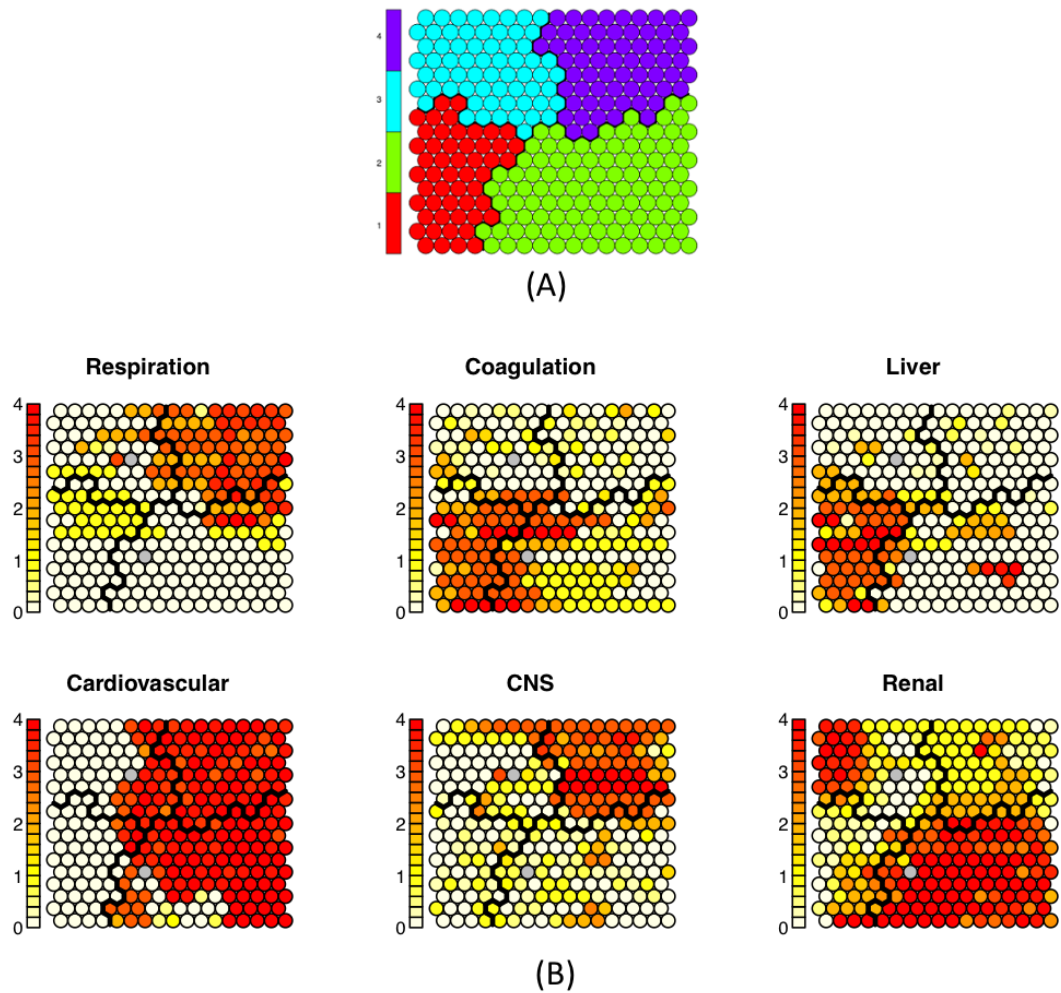

**Figure 2:** (A) The resulting Kohonen Self-Organizing Map depicting overall clusters, colour-coded and separated by black lines. Each SOM node contains 5–35 ICU stays extremely similar to each other.. (d) Shows those same clusters, but with depictions of individual SOFA subscores of the nodes within each cluster. Within each node, the given value is represented by the darkness of the colour in the node. Each node is shaded from white to red, where darker colours represent higher average values (i.e., higher SOFA subscores) among the ICU stays in the given node. The distribution of the SOFA subcomponents shows clear distinctions among the four clusters. The patterns visible in

(B) suggest that the four clusters represent: 1. Liver disease, 2. cardiogenic dysfunction with elevated creatinine, 3. minimal organ dysfunction, and 4. cardiogenic dysfunction with hypoxemia and altered mental status.

**Sepsis Identification**

Using the 13,728 ICU records (9,472 patients without sepsis and 4,256 with sepsis), we performed two binary classification tasks with sepsis diagnosis as outcome. The first task (Class-NoSub) uses the feature set selected using the entire septic population (*FSS*), while the second (Class-Sub) uses the combined feature sets discovered from the septic subpopulations (*FS1-FS4*).

Each of the classification tasks was implemented using three different algorithms via the R package MLR[36]: Random Forest, eXtreme Gradient Boosting (XGBoost) and Support Vector Machines (SVM) using a Gaussian kernel to accommodate non-linearity in the feature space. We optimised each classifier's parameters through a bootstrapped grid search over the respective classifier's' hyperparameter space. All classifiers were trained over 1,000 iterations of a 10-fold cross-validation.

**RESULTS**

The clustering procedure uncovered four subpopulations with distinct organ dysfunction patterns in septic patients in the MIMICIII database (Figure 2c). Further characteristics of the four clusters are given in Table 1. The clusters have been found to represent (1) liver disease, (2) cardiogenic and renal dysfunction, (3) minimal organ dysfunction, and (4) cardiogenic dysfunction with hypoxemia and altered mental status. The subpopulations identified are mostly independent of the origins of sepsis, with etiology being widely distributed across clusters. All predictors showed significant difference across clusters (all p-values < 0.01).

|  | Liver Disease | Cardio & Renal Dysfunction | Minimal Organ Dysfunction | Cardio, Resp. & CNS Dysfunction | All Septic Population |
|---|---|---|---|---|---|
| **Count** | 983 | 1441 | 953 | 879 | 4256 |
| **Age** | 62 (51,72) | 69 (59,79.1) | 69 (57,81) | 76 (68,86) | 68 (58,81) |
| **Female** | 43% (412) | 39% (533) | 39% (371) | 46% (404) | 40% (1720) |
| **Total SOFA (IQR)** | 11.2 (8.5,12.3) | 11.8 (11.1,16.9) | 6.7 (3.4,8.2) | 12.1 (9.5,15.7) | 11.1 (4-15) |
| **Pulmonary SOFA** | 0.3 (0,.6) | 0.8 (0,1.2) | 1.1 (0,2.4) | 3.2 (2,4) | 1.2 (0,3) |
| **Coagulation SOFA** | 3.1 (1.8,4) | 1.5 (0,2.1) | 1 (0,2.1) | 0.5 (0,1) | 1.53 (0,2.9) |
| **Hepatic SOFA** | 3.3 (2.5,4) | 0.7 (0,1.1) | 0.7 (0,1.1) | 0.4 (0,1) | 0.9 (0,1.2) |
| **Cardio SOFA** | 0.8 (0,2) | 3.5 (3,4) | 1.8 (0,4) | 3.8 (3,4) | 2.9 (0,4) |
| **CNS SOFA** | 0.3 (0,1) (0- | 0.6 (0,1) (0.6- | 0.9 (0,2) (0.8- | 3 (0,4) (1.8-4) | 0.7 (0,2) (0- |

|  |  | 1.2) | 0.9) | 1.1) |  | 3.4) |
| --- | --- | --- | --- | --- | --- | --- |
| **Renal SOFA** |  | 1.7 (0,2) | 3.1 (1,4) | 1.4 (0,2.1) | 1.3 (0,2) | 2.6 (1,3.1) |
| **Comorbidity Elixhauser Index (IQR) (95% CI)** |  | 12.4 (5,18.9) | 16 (8.7-21.8) | 7.3 (5,16) | 14 (6,19) | 14.1 (6.6,21) |
| **30-day Mortality** |  | 28% (220) | 55% (724) | 25% (238) | 37% (325) | 35% (1507) |
| **Length of stay (IQR) (95% CI)** |  | 4.3 (1.1-4.8) | 5.6 (2-6.1) | 3.8 (1.8-4.2) | 8.5 (3.2-11.9) | 5.4 (1.8-6.2) |
| **Etiology** |  |  |  |  |  |  |
| Pneumonia |  | 30% (295) | 31% (447) | 35% (334) | 43% (379) | 38% (1617) |
| Urinary Tract |  | 21% (206) | 32% (461) | 27% (257) | 13% (115) | 23% (979) |
| Abdominal |  | 9% (88) | 8% (115) | 7% (67) | 11% (98) | 8% (341) |
| Biliary |  | 12% (118) | 2% (29) | 2% (19) | 2% (19) | 3% (128) |
| Soft Tissue |  | 12% (118) | 15% (216) | 11% (104) | 9% (80) | 11% (468) |
| Other |  | 16% (158) | 12% (173) | 18% (172) | 21% (186) | 17% (723) |

**Table 1:** Cluster Descriptive Statistics of the clusters formed using 31% of the records with an ICD-9 diagnosis of sepsis. Comparing the central tendencies using the Kruskal–Wallis test and found that the clusters significantly differ in length of stay (p-value < 0.01), Elixhauser-Quan comorbidity score (p-value < 0.01) and in-hospital 30-day mortality (p-value <0.01). Etiology values showed no significant differences (p-value ≥ 0.01).

The identified feature sets *FS1-FS4* exhibit direct relevance to the types of organ dysfunction prevalent in the corresponding cluster (detailed in Figure 3a-e). In contrast, the feature set *FSS* includes a subset of the combined *FS1-FS4*, but largely consists of general signs of deterioration (e.g. systolic blood pressure, white blood cell count), which have been found to be good indicators of acuity, but non-discriminative between septic and non-septic patients[37]. The feature selection performance (Figure 3f) indicates the promising contribution of the respective high-resolution feature sets in reducing the error rates of their respective classifiers compared to the low-resolution features.

The classification results comparing Class-Sub and Class-NoSub using RFs, SVMs and XGBoost are presented in Table 2. As the table shows, while XGBoost outperformed other algorithms in all cases, all classifiers showed notably improved performance in the Class-Sub classification task (using the combined feature set *FS1-FS4*). The improved performance is especially pronounced in the specificity of predictions, in which classification without subpopulation features consistently underperforms by not distinguishing septic patients from those with inflammations and comorbidities; a general bottleneck in ML sepsis prediction[11].

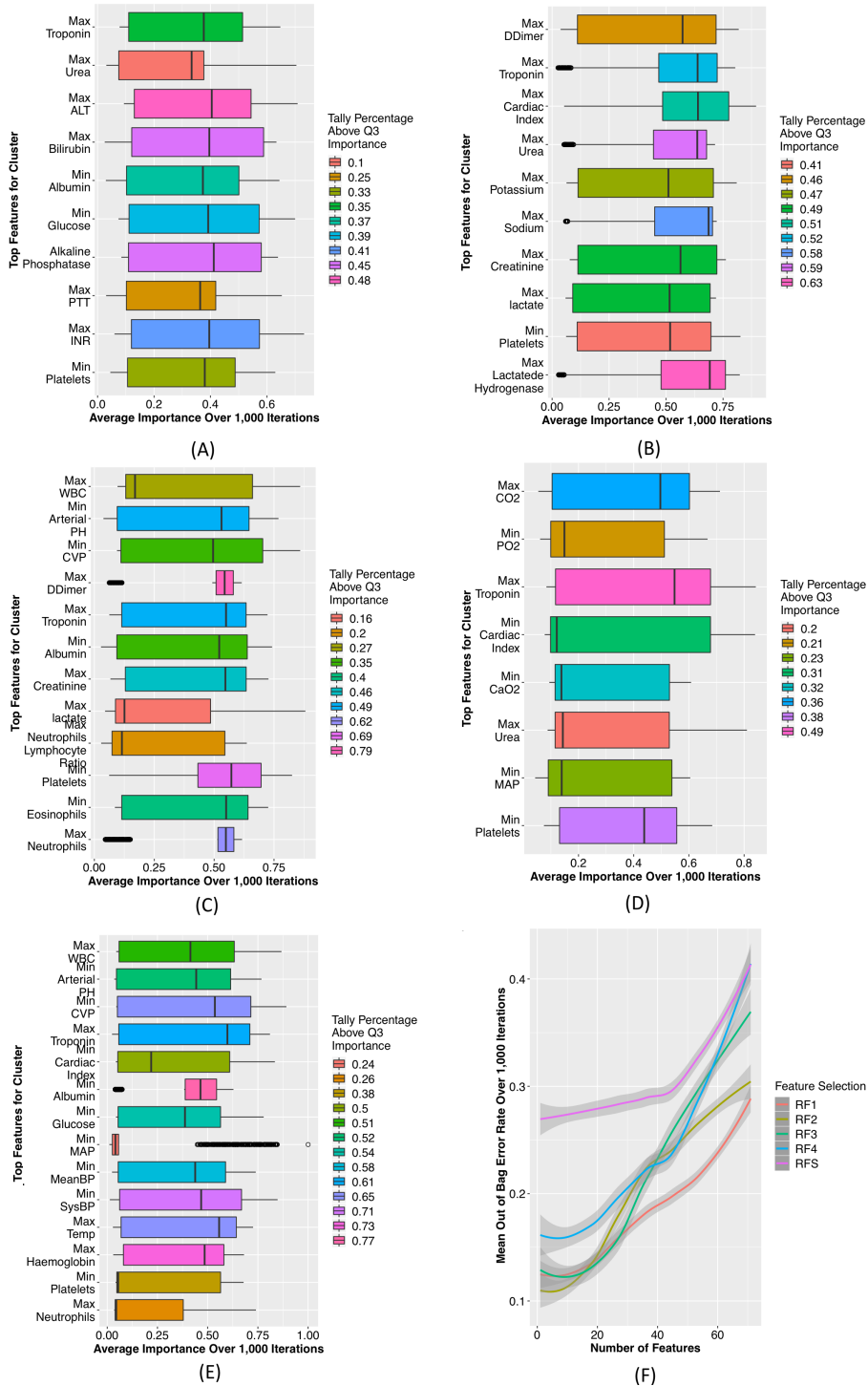

**Figure 3:** Feature selection results: (A) *FS1* comprises high-profile hepatic and coagulation indicators (including bilirubin, Alanine transaminase (ALT) and the International Normalised Ratio (INR)), (B) *FS2* features mainly comprise cardiovascular (troponin, lactate, d-dimer) and renal (creatinine, potassium and urea) indicators, (C) *FS3* features are relevant to all SOFA subcomponents, reflecting the heterogeneity of the minimal organ dysfunction, (D) *FS4* features are of respiratory ($CO_2$, $CaO_2$ and $PO_2$),

and cardiovascular relevance (including mean arterial pressure (MAP)). (E) In contrast, *FSS* includes a portion *FS1-FS4*, but largely consists of general signs of deterioration (e.g. systolic blood pressure, temperature, white blood cell count (WBC)), which have been found to be good indicators of mortality, but non-discriminative between septic and non-septic patients[31]. (F) shows the out-bag-error rates over 1,000 iterations of bootstrapped feature selection.

| Classifier | Model | Sensitivity (95% CI) | Specificity (95% CI) | AUC | PLR (95% CI) | NLR (95% CI) |
|---|---|---|---|---|---|---|
| **RF** | ClassNoSub | 0.85 (0.81-0.88) | 0.79 (0.74-0.78) | 0.8 | 4 (4.9-4.2) | 0.18 (0.16-0.19) |
| | Class-Sub | 0.94 (0.92-0.95) | 0.93 (0.91-0.95) | 0.94 | 13.4 (20.34-22.29) | 0.06 (0.03-0.08) |
| **XGBoost** | Class-NoSub | 0.87 (0.85-0.91) | 0.81 (0.80-0.4) | 0.84 | 4.6 (4.1-4.9) | 0.16 (0.1-0.21) |
| | Clas-Sub | 0.96 (0.94-0.97) | 0.95 (0.93-0.97) | 0.96 | 19.12 (18.6-46.18) | 0.04 (0.02-0.05) |
| **SVM** | Class-NoSub | 0.86 (0.88-0.91) | 0.83 (0.84-0.86) | 0.84 | 5.1 (4.5 – 6.2) | 0.17 (0.13-0.2) |
| | Class-Sub | 0.95 (0.93-0.96) | 0.93 (0.92-0.95) | 0.94 | 13.6 (11.51 – 16.65) | 0.07 (0.02 – 0.12) |

**Table 2:** Performance metrics for the two classification tasks Class-NoSub (using features selected without distinguishing subpopulations) and Class-Sub (using features selected from the individual subpopulations), using Random Forests (RF), Support Vector Machines (SVM) and Extreme Gradient Boosting (XGBoost). PLR: Positive Likelihood Ratio, NLR: Negative Likelihood Ratio.

**DISCUSSION**

The principle goal of this work is to evaluate the importance of accounting for the heterogeneity of sepsis in improving early diagnosis. Current ML prediction models operate by linking vital signs to sepsis without acknowledging the heterogeneity of the condition. The subpopulations we derived here exhibit distinct patterns of organ dysfunction and outcome distributions, and are mostly independent of severity or etiology of sepsis. Instead, there exist significant correlations in each sepsis subpopulation between specific combinations of vital sign values and sepsis diagnoses. Because the results show the superiority of subpopulation-specific features in discriminating septic from non-septic patients, further research in early sepsis diagnosis can improve prediction quality by focusing on the recognition of the patterns of vital sign changes of specific relevance to sepsis subpopulations as opposed to the generic features currently used in ML prediction models.

It is important to place our findings in the context of the goal of this work: the clusters identified are representative of ICU admissions in a single hospital and are only used to support our central hypothesis, and not to suggest the discovery of new sepsis subtypes. Similarly, although the experiments show an improved predictive power, they do not constitute a new sepsis prediction algorithm; doing the latter requires examining the temporal evolution of predictors, which we chose to exclude from our evaluations as it adds a level of complexity not needed to evaluate our premise. Nevertheless, by choosing the most abnormal value for each variable, we single out the maximal physiological derangement connecting the given feature to the outcome. In addition, our findings are not intended to discover new rules for manual scoring systems, but to direct future development of EHR-integrated decision support tools. Besides early diagnosis, the direction provided by the subpopulations can be further developed to issue justifications and explanations to clinicians, including embedded visualisations.

The particular nature of MIMICIII renders the results specific to sepsis developed in the ICU. While early diagnosis of sepsis in ICU settings is an important problem[38], our findings have not been verified non-ICU wards where more than 50% of sepsis cases are identified. However, by obtaining aggregate values of patient vitals over the first 24 hours of ICU admission, we aimed to obtain a clinical picture similar to the worst value obtained over the last hours in hospital wards. Nonetheless, our on-going work targets the generalisation of the model to non-ICU settings.

Finally, suspicion of sepsis, as defined by the co-occurrence of culture and antibiotic, is systematically underrepresented in a large portion of MIMICIII records[15]. Therefore, although ICD codes do not constitute a perfect representation of the true incidence of sepsis in EHR records[39], we used them in conjunction with culture results to identify septic records.

**CONCLUSIONS**
Current sepsis prediction tools are yet to take into account the known heterogeneity of the condition. Our study found that accounting for clinically meaningful subpopulations within a large ICU sepsis cohort highly improved predictive power. The findings of our work can guide future sepsis prediction towards more accurate and explainable models. However, more multi-centre studies are warranted on the etiological and vital variations within septic patients, and for evaluation.

**FUNDING STATEMENT**

This work was supported by the following sources:

1. ZI and RD are supported by the National Institute for Health Research (NIHR) Biomedical Research Centre at South London and Maudsley NHS Foundation Trust and King's College London.
2. RD is supported by the Health Data Research UK, which is funded by the UK Medical Research Council, Engineering and Physical Sciences Research Council, Economic and Social Research Council, Department of Health and Social Care (England), Chief Scientist Office of the Scottish Government Health and Social Care Directorates, Health and Social Care Research and Development Division (Welsh Government), Public Health Agency (Northern Ireland), British Heart Foundation and Wellcome Trust.
3. ZI and RD are supported by the National Institute for Health Research University College London Hospitals Biomedical Research Centre.
4. HW is funded by MRC grants ( MR/S004149/1 and MC_PC_18029).

**COMPETING INTERESTS STATEMENT**

The authors have no competing interests to declare.

**CONTRIBUTIONSHIP STATEMENT**

AA and ZI conceived the research idea. ZI and HW designed the machine learning pipeline. LS performed auxiliary analyses to isolate the machine learning models to be used in this work. ZI and AA performed data acquisition. AH performed manual verification of the septic cohort from the MIMICIII clinical admission notes. All authors made substantial contributions to the analysis and interpretation of the results. ZI drafted the manuscript with all authors contributing to the critical revision for important intellectual content and addressing the reviewers' comments. All authors agree to be accountable for all aspects of the work in ensuring that questions related to the accuracy or integrity of any part of the work are appropriately investigated and resolved.

# APPENDIX A: SEQUENTIAL ORGAN FAILURE (SOFA) SUBSCORES

Respiratory system

| PaO2/FiO2 (mmHg) | SOFA Score |
|---|---|
| ≥ 400 | 0 |
| 300-399 | +1 |
| 200-299 | +2 |
| 100-199 and mechanically ventilated | +3 |
| 0-99 and mechanically ventilated | +4 |

Nervous system

| Glasgow Coma Scale | SOFA Score |
|---|---|
| 15 | 0 |
| 13-14 | +1 |
| 10-12 | +2 |
| 6-9 | +3 |
| 3-5 | +4 |

Cardiovascular System

| Mean Arterial Pressure OR Administration of Vasopressors Required | SOFA Score |
|---|---|
| MAP ≥ 70 mmHg | 0 |
| MAP < 70 mmHg | +1 |
| dopamine ≤ 5 μg/kg/min or dobutamine (any dose) | +2 |
| dopamine > 5 μg/kg/min OR epinephrine 0.1 ≤ μg/kg/min OR norepinephrine ≤ 0.1 μg/kg/min | +3 |
| dopamine > 15 μg/kg/min OR epinephrine > 0.1 μg/kg/min OR norepinephrine > 0.1 μg/kg/min | +4 |

Liver

| Bilirubin (mg/dl) (μmol/L) | SOFA Score |
|---|---|
| < 1.2 (< 20) | 0 |
| 1.2–1.9 (20-32) | +1 |
| 2.0–5.9 [33-101] | +2 |
| 6.0–11.9 (102-204) | +3 |
| > 12.0 (>204) | +4 |

Coagulation

| Platelets×10³/μl | SOFA Score |
|---|---|
| ≥150 | 0 |
| 100- 149 | +1 |
| 50-99 | +2 |
| 20-49 | +3 |
| 0-19 | +4 |

Renal

| Creatinine (mg/dl) (μmol/L) (or urine output) | SOFA Score |
|---|---|
| < 1.2 (< 110) | 0 |
| 1.2–1.9 (110-170) | +1 |
| 2.0–3.4 (171-299) | +2 |
| 3.5–4.9 (300-440) (or < 500 ml/d) | +3 |
| > 5.0 (> 440) (or < 200 ml/d) | +4 |

| Creatinine (mg/dl) (μmol/L) (or urine output) | SOFA Score |
|---|---|
| < 1.2 (< 110) | 0 |
| 1.2–1.9 (110-170) | +1 |
| 2.0–3.4 (171-299) | +2 |
| 3.5–4.9 (300-440) (or < 500 ml/d) | +3 |
| > 5.0 (> 440) (or < 200 ml/d) | +4 |

# APPENDIX B: FEATURES AND AGGREGATION PROCEDURES

| Variable | Features Extracted |
|---|---|
| Troponin, Heartrate, Lactatede Hydrogenase, Creatinine, DDimer, INR, PTT, Fibrinogen, Bilirubin, AST, ALT, Urea, DeltaCO2, Lactate | Maximum |
| Haemoglobin, CAO2, Heartrate, Sodium, Bicarbonates, Systolic BP, Diastolic BP, Mean BP, Mean Arterial Pressure, Stroke Index, Peripheral Saturation, PF Ratio, platelets, Albumin | Minimum |
| Temperature, RespiratoryRate, CardiacIndex, SystemicVascularResistanceIndex, CentralVeneousPressure, WhiteBloodcellCount, Neutrophils, Eosinophils, Lymphocytes, AtypicalLeukocytes, Bandforms, Glucose, Arterialph, Sodium, Potassium, Chloride, Magnesium, Neutrophils Lymphocyte Ratio | Minimum and Maximum |
| Alkaline Phosphatase | Sum |